\documentclass[12pt,aps]{revtex4}
\usepackage{amsmath,amssymb,graphicx,epsfig}
\begin{document}

%%%%%%%%%%%%%%%%%%%%%%%%%%%%%%%%%%%%%%%%%
\title{Entanglement Induced Fluctuations of Cold Bosons}
\author{Ram Brustein, Amos Yarom}
\affiliation{ Department of Physics, Ben-Gurion University, Beer-Sheva 84105, Israel \\
    {\rm E-mail:}
    {\tt ramyb@bgumail.bgu.ac.il, yarom@bgumail.bgu.ac.il } }

%%% ----------------------------------------------------------------------

\begin{abstract}

We show that due to entanglement, quantum fluctuations may differ significantly
from statistical fluctuations.
We calculate quantum fluctuations of the
particle number and of the energy in a sub-volume of a system of bosons in a pure
state, and briefly discuss the possibility of measuring them. We find that energy
fluctuations have a non-extensive nature.
\end{abstract}

\pacs{ }

\maketitle
%%% ----------------------------------------------I------------------------
\section{Introduction and summary}
In 1964 Bell showed, using his famous inequality, that a system
prepared in an entangled state can not be described by local,
classical physics \cite{Bell}. Since then, different kinds of
`Bell-type-inequalities' have been found and studied. The special
nature of an entangled state has lead to many developments in
quantum teleportation, quantum games, and even in quantum black hole
physics \cite{'tHooft:1985}. In this paper we show that entanglement
is also relevant to the nature of quantum fluctuations (See, \cite{buttiker,castin} for interesting recent discussions of the subject.)

We discuss a non-relativistic many particle system in a volume
$\Omega$. We divide the whole volume $\Omega$, into some
sub-volume $V$, and its complement $\widehat{V}$. The system is
initially prepared in a pure state $\langle \vec{x} |\psi\rangle$ defined in $\Omega$. We consider states $|\psi\rangle$ which are entangled with respect to the Hilbert spaces of $V$ and $\widehat{V}$. This means that $| \psi \rangle$ can not be
brought into a product form $| \psi \rangle = | \psi \rangle_1
\otimes | \psi \rangle_2$ in terms of a pure state $| \psi \rangle_1$
that belongs to the Hilbert space of $V$, and another pure state $|
\psi \rangle_2$ that belongs to the Hilbert space of $\widehat{V}$.
Consider an operator $O^V$ that is restricted to the sub-volume $V$.
We shall show that the quantum fluctuations of $O^V$: $ (\Delta O^V)^2= \langle \left(O^V\right)^2\rangle -\left(\langle O^V\rangle\right)^2$, behave differently from standard statistical fluctuations due to the entangled nature of the initial pure state $|\psi\rangle$.

Specifically, we consider energy fluctuations in a sub-volume of a system of cold
bosons of mass $m$ in  its ground state  , i.e., a Bose-Einstein condensate (BEC). The BEC is assumed to be in a periodic box of size $L^d$ ($d$ being the dimension of space.) Had the
fluctuations been thermal, we would have expected them to be extensive, and so, to
scale as the volume: $\text{Trace} (\rho (\Delta E^V)^2) \propto V$. Instead, we find a different behaviour, for example, the energy fluctuations in a
sub-volume of size $a^d$ scale as the surface area of the sub-volume
\begin{equation}
 \label{efluctuationsA}
    \Delta E^2=\frac{4 n d \pi^4 \hbar^4 Q^2 k^2 }{m^2 L^{3+d}}a^{d-1}+\cdots.
\end{equation}
Here $\pi^2 Q^2$ is a high momentum (UV) cutoff that needs to be introduced in
order to render the energy fluctuations finite. The UV cutoff is determined by the
effective width of the boundary that separates the sub-volume from the rest of the
system. The dots represent sub-leading terms in an expansion in $1/Q$.
We shall discuss this cutoff shortly.

The area-scaling is not a special feature of the energy fluctuations in the ground state. General
arguments \cite{AreaScaling} show that it occurs for a large class of operators
whose two-point function is short ranged. It has also been shown that the area
scaling is robust to changes in the cutoff scheme and does not depend sensitively
on the nature of the pure state.

The energy fluctuations exhibit a power law dependence on the high momentum cutoff.
This may seem strange and counter intuitive. One may ask why should a macroscopic quantity
depend so strongly on the details of the physics that are relevant only to very
short distances? Our explanation for these power law divergences relies on the uncertainty principle~\footnote{We thank Yvan Castin for suggesting this explanation.};
a measurement of the momentum fluctuations of a localized particle is formally divergent. Smearing the position of the particle will render the momentum fluctuations finite and inversely dependent on the smearing scale.
The divergence in the energy fluctuations is of similar origin: measuring the energy fluctuations in a volume with a sharp boundary yields an infinite result. Introducing a high momentum cutoff is equivalent to smearing the boundary of the sub-volume. The width of the boundary is inversely dependent on the cutoff scale.

A measurement of energy fluctuations in the laboratory may shed some light on these
issues. Experiments in BEC type systems seem to offer a unique opportunity for
measuring entanglement induced fluctuations. To perform such an experiment one
needs to be able to make repetitive measurements of a particular sub-volume of the
condensate, and prepare the condensate in the same initial state for each
measurement. Alternately, one may make measurements of several sub-volumes in the
same condensate using tomographic methods. A verification of the explicit
dependance on the high momentum cutoff of energy fluctuations is not only
interesting in its own right, but may also have implications concerning Unruh
radiation and black hole physics \cite{TandA}.
In the BEC context one may guess that the UV momentum cutoff is inversely
proportional to the resolution of the measuring device.
If so, our result in eq.(\ref{efluctuationsA}) implies that a
better detector resolution  will lead to larger energy fluctuations.

Much has been said about the area law of entanglement entropy (see for example \cite{Plenio,yurtsever}) and black holes \cite{Srednicki,Bombelli,Kabat,KabatStrassler}. To understand the possible relationship of entanglement induced fluctuations to
black hole physics it is useful to consider as an example the vacuum state of some
field(s) in the space-time of an eternal black hole. An observer outside the black
hole will make measurements that are restricted to the exterior of the event
horizon. Let us take for example the measurement of energy fluctuations by this
observer. They are divergent, depend on a power of the high momentum cutoff, as is
the entanglement entropy, and scale as the surface area of the horizon. The
divergence originates from distance scales which are very close to the horizon.
Technically they arise from the divergence of the two-point function of the energy
density at short distance scales. This is the same mechanism which generates the
divergences in the case of cold bosons.
The understanding that we have gained on the nature of the divergences
in the non-relativistic setting  provides a new perspective on the nature
of the same divergences that appear in the calculation of energy fluctuations in black hole spacetimes.

In addition to the energy fluctuations, one may also consider fluctuations in the
particle number (see \cite{pitaevskii} for a recent discussion.) We give an analytic expression for the fluctuations of the particle
number in various potentials in section \ref{S:pnumber}. The particle number
fluctuations are different from the energy fluctuations in that they do not depend
on the high momentum cutoff. This is because the two point function of the number
density operator does not diverge strongly enough at short distances
\cite{AreaScaling}. Yet, we show that the quantum correlations in the initial state
generate fluctuations which are different from the ones that are expected in a
classical setup.

%%% ------------------------------------------------------------------------
\section{General setup}

We shall describe the general method of evaluating fluctuations of
operators restricted to subvolumes in the case that the whole system
is in a pure state $|\psi\rangle$. We will consider the case that the pure state has a fixed number of particles and fixed momentum.  We discuss systems whose Hamiltonian is time-independent and have some fixed boundary conditions (BC), for example, periodic or Dirichlet. For the general discussion we do not specify the BC, whose significance will be discussed later when we consider in detail fluctuations of two specific operators: the particle number operator, and the energy operator.

We start with a general operator density $\mathcal{O}(\vec{x})$. It can be expressed in terms of the particle field $\psi(\vec{x},t)=\sum_i a_i u_i(\vec{x}) e^{-\imath E_i t}$ as follows,
$ \mathcal{O}(\vec{x})=\psi^{\dagger}(\vec{x},t)
\widehat{\mathcal{O}}
                         \psi(\vec{x},t).
 $
The functions $u_i$ are a basis of eigenfuctions of the Hamiltonian.
For example, the energy operator is given by
$\widehat{E}=-\frac{1}{2m}\nabla^2 + V$, and the particle number operator by
$\widehat{N}=I$.

We consider a pure state $|\psi\rangle$ with $n$ particles of
momentum $k$. Evaluating the operator $O^V=\int_V
\mathcal{O}(\vec{x}) dV$ in this state gives
\begin{align}
    \langle \psi | O^V | \psi \rangle
    &=
        \langle \psi | \int \sum_{l,p} a^{\dagger}_l a_p
    u_l^{\star}(\vec{x}) \widehat{\mathcal{O}} u_p(\vec{x}) dV | \psi \rangle \nonumber \\
    &=
        \int
        u_{k}^{\star}(\vec{x}) \widehat{\mathcal{O}} u_{k}(\vec{x}) dV
    \equiv n \widehat{O}_{k,k}.
\end{align}
The fluctuations of the operator $O^{V}$, $\langle\psi | (\Delta
O^V)^2 |\psi \rangle= \langle\psi |  (O^V)^2 | \psi\rangle -
(\langle \psi |O^V | \psi\rangle)^2$, are given by
\begin{align}
 \label{fluctuations}
    \langle \psi | (\Delta O^V)^2 | \psi \rangle
    &=
    n \sum_{s} \widehat{O}_{k s} \widehat{O}_{s k}-
    n \widehat{O}_{k k}^{2} \nonumber\\
    &=
    n \sum_{s \neq k} \widehat{O}_{k s} \widehat{O}_{s k}.
\end{align}

One may generalize this result to a system with $T$ particle modes,
such that there are $n_i$ ($1 \le i \le T$) particles with quantum
numbers $k_i$, $|\psi\rangle = \Pi_{i=1}^{T}
\frac{(a^{\dagger}_{k_{i}})^{n_i}}{\sqrt{n_{i}!}}|0\rangle$.
%%CHANGE
The most general state will be a linear superposition of such multi
particle states.
%%ENDCHANGE
In this case we have
\begin{align}
    \langle \psi | O^V | \psi \rangle
    &=
    \sum_{i=1}^{T} n_{i} \widehat{O}_{k_{i},k_{i}},\\
\intertext{and}
    \langle \psi | (O^V)^2 | \psi \rangle
    &=
    \sum_{s \neq i} n_{i} \hat{O}_{k_{i} s} \hat{O}_{s k_{i}}+
    \sum_{e \neq i} n_{i} n_{e} \hat{O}_{k_{i} k_{e}} \hat{O}_{k_{e} k_{i}}.
\end{align}

In order to study the properties of the fluctuations we need to
evaluate the matrices $\widehat{O}_{k s}=\int u_{k}^{\star}(\vec{x})
\widehat{\mathcal{O}} u_{s}(\vec{x}) dV$. We shall consider two
special cases: the number of particles operator and the energy
operator. For the number operator we have $\widehat{N}=I$. Then
\begin{align}
    \sum_{s}  \widehat{N}_{ks} \widehat{N}_{sl} &=
    \sum_{s} \int_{V} u_{k}^{\star}(\vec{x}) u_{s}(\vec{x}) dV_x
             \int_{V} u_{s}^{\star}(\vec{y}) u_{l}(\vec{y}) dV_y \nonumber \\
    &=\int_{V} u_{k}^{\star}(\vec{x}) \delta(\vec{x}-\vec{y}) u_{l}(\vec{y})
    dV_x dV_y = \widehat{N}_{kl}.
\end{align}
So for one particle mode we get
\begin{equation}
 \label{nfluctuations}
  (\Delta N^{V})^{2} = n \widehat{N}_{kk} - n
    \widehat{N}_{kk}^{2} = \langle N^{V} \rangle (1-\widehat{N}_{kk}).
\end{equation}

For the energy operator, we have:
\begin{align}
\label{efluctuations}
    \widehat{E}_{ks} &= \int_{V} u^{\star}_{k}(\vec{x}) \widehat{H} u_{s}(\vec{x}) dV
    \nonumber \\
    &= E_{s} \widehat N_{ks}
\end{align}
So that for the case of one particle mode we get:
\begin{equation}
\label{E:energy_fluc}
 (\Delta E^{V})^{2} = n \sum_{s \neq k} E_{k} E_{s}
     \widehat{N}_{ks} \widehat{N}_{sk}.
\end{equation}

%-----------------------------------------------------------------------------------
\section{Relative particle number fluctuations}
\label{S:pnumber}
We proceed to consider in detail the relative particle number
fluctuations in the case that the initial state contains a single
particle mode: $(\Delta N^{V})^{2} / \langle N^{V} \rangle = 1-\widehat{N}_{kk}$.

\subsection{Free particles in a box}

The simplest case to consider is a one dimensional box of size $L$,
with periodic BC. The solutions to the
Schr\"{o}dinger equation are of the form
 $
    u_n(x) = \frac{1}{L^{1/2}} e^{2\pi\imath n x/L},
 $
with $n \in \mathbb{Z}$. One obtains $N_{n,n}= \frac{a}{L}$. We also give
the result for the case  $m \neq n$ for future reference,
\begin{align}
\label{E:N_for_pbox}
    \widehat{N}_{n,m} & = 1/L \int_0^a e^{-2\pi\imath n x/L} e^{2 \pi \imath mx/L}dx\\
\notag
                 & = \frac{1}{\pi (m-n)} e^{\pi \imath (m-n)a/L} \sin(\pi (m-n)
                 a/L).
\end{align}
One then has $(\Delta N_V) ^2/\langle N_V \rangle = 1
- a/L$. Generalizing to d-dimensions, we get
\begin{equation}
\label{deltand1}
(\Delta N_V) ^2/\langle N_V \rangle = 1 - (a/L)^{d}.
\end{equation}

\begin{figure}[btp]
\begin{center}
\includegraphics{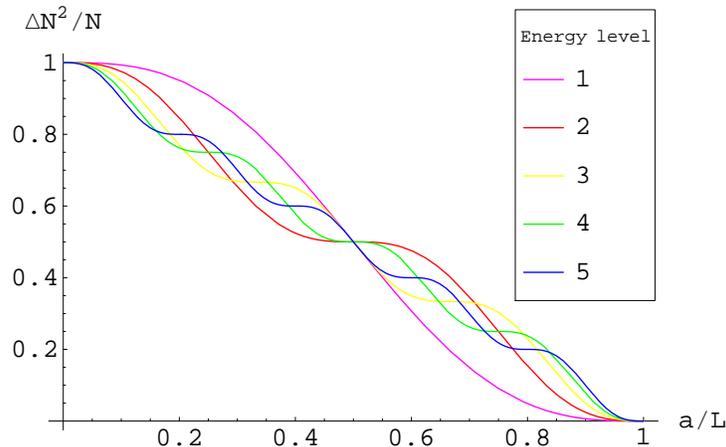}
\caption{ \label{figure1} Shown is a graph of the relative particle number
fluctuations $(\Delta N_V) ^2/\langle N_V\rangle$ in a sub-volume of
size $a$ of a one dimensional box of size $L$ with Dirichlet boundary conditions
versus the relative size of the sub-volume $a/L$. Each color corresponds to a
different initial state as specified by the legend.}
\end{center}
\end{figure}

The eigenstates for free particles in a box with Dirichlet BC (Dirichlet box) are:
\[
    u_n(x) = \sqrt{2/L} \sin(n \pi x/L),
\]
and so
\begin{equation}
    \widehat{N}_{n,m} = \frac{1}{\pi} \left(
                            \frac{\sin((m-n)\pi a/L)}{m-n} -
                            \frac{\sin((m+n)\pi a/L)}{m+n}
                         \right).
\end{equation}
The diagonal matrix elements are
\[
    \widehat{N}_{k,k}=\frac{a}{L}-\frac{\sin\left(2\pi k \frac{a}{L}\right)}{2\pi k }.
\]
In d-dimensions one has
\begin{equation}
\label{deltandd}
(\Delta N_V) ^2/\langle N_V \rangle = 1
-\left(\frac{a}{L} - \frac{\sin\left(2\pi k \frac{a}{L}\right)}{2\pi
k}\right)^{d}.
\end{equation}
The first five levels for a one dimensional system are shown in
Figure~\ref{figure1}. The results for the Dirichlet box are qualitatively similar
to the results of the periodic box, the main difference being the oscillations that are superimposed on the factors of $a/L$. These oscillations are purely quantum and do not appear in an equivalent microcanonical ensemble as we show next.

We would like to compare our results in equations (\ref{deltand1}) and (\ref{deltandd}) to their classical analogs. We show that the case of the periodic box gives identical results to the case of a classical uniform distribution of particles in a box. The case of Dirichlet box is different from the classical example and exhibits interesting quantum oscillations.

Let us consider a classical system with $n$ particles, uniformly distributed  in a box of size $L$ and calculate $(\Delta N_V) ^2$ and $\langle N_V \rangle$. The probability for finding a particle in a given position is : $P(x)dx = 1/L\,dx$. Therefore the probability of finding a particle in a
region of size $a<L$ is $p=a/L$. The probability of not finding it
there is $q=1-a/L$. If there are $N$ particles and they are
distributed so that each one has a uniform probability to be found
in a given position, then the probability of finding exactly $N_a$
particles inside a region of size $a$ is binomial
\begin{equation}
    P(N_a | N) = \left (
                    \begin{array}{c}
                        N \\
                        N_a \\
                    \end{array}
                 \right )
                 p^{N_a}q^{N-N_a}.
\end{equation}
Therefore the mean number of particles is $\langle N_a \rangle = N
p$, and the variance is $ \Delta N_a^2 = N p q$. So
that
\begin{equation}
\label{deltanclass}
    \frac{ \Delta N_a^2 }{\langle N_a \rangle} = q
    =1-\frac{a}{L}.
\end{equation}
This result generalizes in an obvious way to $d$-dimensions $\frac{ \Delta N_a^2 }{\langle N_a \rangle} = 1-\left(\frac{a}{L}\right)^d$, which is exactly the result in eq.(\ref{deltand1}). The behavior exhibited in eq.(\ref{deltanclass}) is quite different from the one exhibited in eq.(\ref{deltandd}) for the Dirichlet BC case where the correlations may induce a sinusoidal variation in the
fluctuations (see Figure~\ref{figure1}.)

\subsection{The harmonic oscillator}

The BEC condensate may be described as a collection of bosonic particles in a
harmonic oscillator potential (see \cite{BEC} for a review), so it is interesting
to consider the particle number fluctuations in such a setup. The eigenstates for a
one dimensional harmonic oscillator are given by
\[
    u_n(x) = \frac{1}{\sqrt{n!}}2^{-n/2}\left(\frac{\eta}{\sqrt{\pi}}\right)^{1/2}
             H_n \left(\eta x\right)
             e^{-\frac{\eta^2 x^2}{2}},
\]
where we have defined $\eta = \sqrt{m \omega / \hbar}$. Therefore, if our
sub-volume is the region $a < x < b$, then for $m \neq n$ we have:
\begin{equation}
\notag
    \widehat{N}_{n,m}=
\label{E:Nnm_Harmonic_1}
    \frac{\sqrt{n!m!}}{2^{(n+m)/2}\sqrt{\pi}} \sum_{l=0}^{n}
    \frac{2^l}{l!(m-l)!(n-l)!}
    \left(e^{-\eta^2 a^2} H_{m+n-2l-1}(\eta a) - e^{-\eta^2 b^2} H_{m+n-2l-1}(\eta b)
    \right),
\end{equation}
and for $m=n$ we have:
\begin{multline*}
    \widehat{N}_{nn}=
    \frac{n!}{2^n \sqrt{\pi}} \sum_{l=0}^{n-1}
    \frac{2^l}{l!(n-l)!(n-l)!}
    \left(e^{-\eta^2 a^2} H_{m+n-2l-1}(\eta a) - e^{-\eta^2 b^2} H_{m+n-2l-1}(\eta b)
        \right) \\
    +\frac{1}{2}(\text{erf}(\eta b)-\text{erf}(\eta a)).
\end{multline*}

\begin{figure}[btp]
\label{figharmonic}
\begin{center}
\includegraphics{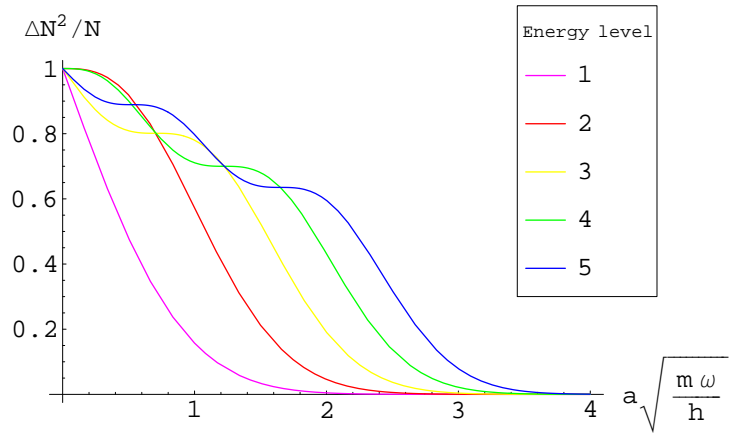}
\caption{\label{figure2} Shown is a graph of the relative particle number
fluctuations $(\Delta N_V) ^2/\langle N_V\rangle$ for the case of a
harmonic oscillator of mass $m$ and frequency $\omega$. The sub-volume is in the
region $-a<x<a$ and the fluctuations are plotted against the relative size of the
sub-volume in units of $\sqrt{m \omega / \hbar}$. Each color corresponds to a
different initial state as specified by the legend.}
\end{center}
\end{figure}

For the case that the sub-volume is symmetric about the minimum of
the potential $-a<x<a$ the expressions simplify and we find that for
the lowest energy state $\langle(\Delta N_V) ^2\rangle/\langle
N_V\rangle = 1- \text{erf}(\eta a)$. This, and a plot of the first
four excited modes (in one dimension) is shown in Figure~\ref{figure2}. For
the lowest level of a d-dimensional symmetric harmonic oscillator we
have $(\Delta N_V) ^2/\langle N_V\rangle = 1-
(\text{erf}(\eta a))^d$.

The particle number fluctuations in the lowest energy state of a
harmonic oscillator have some similarities to the fluctuations of
free particles in a box. The ``size" of the whole system, $L$ in
this case, is approximately $L\sim 1/\eta$ (recall that $\eta =
\sqrt{m \omega / \hbar}$), and for $\eta a<1$, $\text{erf}(\eta
a)\sim 2/\sqrt{\pi} a/L$. It follows that $(\Delta N_V) ^2/\langle N_V\rangle\sim 1- (a/L)^d$. However, in the higher
energy states, the ``size" of the system increases since the wave
function spreads more and more as the energy becomes higher, and the
particle number fluctuations exhibit a different behaviour.

%------------------------------------------------------------------------------------
\section{Energy fluctuations}

The derivation of the result for the energy fluctuations from
eq.(\ref{E:energy_fluc}) is not as tractable as the derivation of
the particle number fluctuations. We evaluate the energy
fluctuations in a cubic sub-volume of a d-dimensional periodic box
and show that the energy fluctuations scale as the surface area of
this sub-volume. It should be clear that result is quite general,
and that similar results should be expected for other setups. There
is also ample additional evidence that this scaling behavior is
quite robust \cite{AreaScaling}.

The expression for energy fluctuations is given in eq.(\ref{E:energy_fluc}). We
start by evaluating $\sum_s E_k E_s \widehat{N}_{ks} \widehat{N}_{sk}$. Using
eq.(\ref{E:N_for_pbox}), we get:
\begin{equation}
 \label{divdeltae}
    \sum_s E_k E_s \widehat{N}_{ks} \widehat{N}_{sk}
    =\frac{16\pi^4
    \hbar^4 k^2}{4 m^2 L^4 \pi^2}
    \sum_{-\infty}^{\infty}
    \frac{s^2}{(k-s)^2}\sin^2(\pi(k-s)\frac{a}{L}).
\end{equation}
This expression is divergent, so we introduce an exponential UV cutoff on the sum  $e^{-\frac{|s|}{\pi^2 Q^2}}$.

As shown in the Appendix, the cutoff introduces a smoothing of the boundary of the sub-volume. So far we have considered a perfectly sharp boundary $0\le x \le a$ and did not encounter any problems. Equivalently, we have defined the sub-volume by integrating with an infinitely  sharp step function $H(x-a)=\begin{cases} 1, & 0\le x \le a  \\ 0, &  x > a \end{cases}$, so $\int\limits_0^a dx = \int\limits_0^L dx H(x-a)$. The smoothing of the boundary is over a region $x\sim L/Q$, and equivalently the smooth step function changes its value from $1$ to $0$ over a range $\sim L/Q$. The specific cutoff scheme determines the detailed behaviour of the smoothed step function over the region where it changes significantly. However, all the cutoff schemes induce very similar qualitative smoothing. The advantage of using exponential cutoff is that the expression for the energy fluctuations can be evaluated analytically with the cutoff.

With the cutoff eq.(\ref{divdeltae}) becomes
\begin{equation}
\label{E:Efluc_Harmonic_1}
    \sum_s E_k E_s
        \widehat{N}_{ks} \widehat{N}_{sk}
    \to
    \frac{E_k^2}{k^2}\frac{1}{\pi^2}
        \sum_{-\infty}^{\infty}
        \frac{s^2}{(k-s)^2}\sin^2(\pi(k-s)\frac{a}{L})e^{-\frac{|s|}{\pi^2 Q^2}}.
\end{equation}
This series may be evaluated analytically. We find
\begin{multline}
\label{E:Efluc_Harmonic_2}
      \pi^2 dE(Q,k,x)\equiv\sum_{\substack{s = -\infty \\ s \neq k}}^{\infty}
        \frac{s^2}{(k-s)^2}\sin^2(\pi(k-s)x)e^{-\frac{|s|}{\pi^2 Q^2}}
   =\\
        -e^{-\frac{k}{\pi^2 Q^2}}\frac{\coth\left(\frac{1}{2\pi^2 Q^2}\right) \sin^2(\pi x)}
              {\cos(2\pi x)-\cosh\left(\frac{1}{\pi^2 Q^2}\right)}
        +e^{-\frac{k}{\pi^2 Q^2}}\frac{1}{2}
              (2\text{Li}_2(e^{-\frac{1}{\pi^2 Q^2}})-
                \text{Li}_2(e^{-\frac{1}{\pi^2 Q^2}-2\imath\pi x})-
                \text{Li}_2(e^{-\frac{1}{\pi^2 Q^2}+2\imath\pi x})),
\end{multline}
where $x=a/L$, and $\text{Li}_2$ is the dilogarithm function.

\begin{figure}[btp]
\label{figenergy1}
\begin{center}
\includegraphics{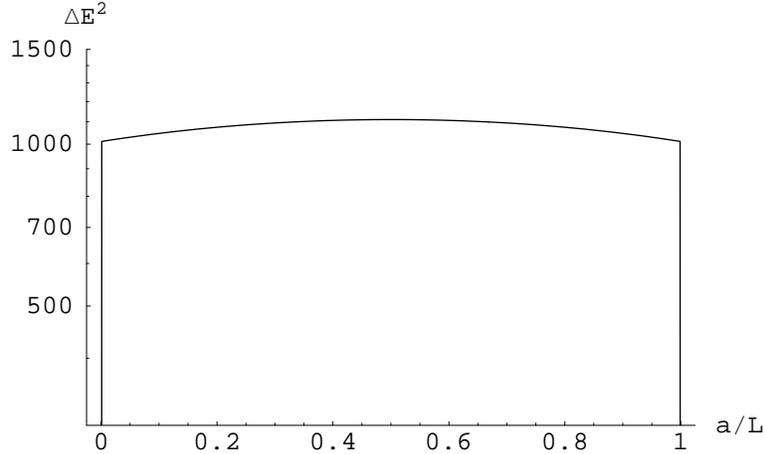}
\caption{\label{figure3} Logarithmic plot of energy fluctuations versus $a/L$ for a
one dimensional box with periodic boundary conditions. Here, the UV cutoff $\pi^2
Q^2$ is 10000, the state is the 20th excited state ($k=20$), and the mass has been
chosen such that $\frac{E_k^2}{\pi^2 k^2}=1$.}
\end{center}
\end{figure}

Therefore, the energy fluctuations for $n$ particles in one dimension are given by
\begin{equation}
\label{E:harmonic3}
    \Delta E^2=n\frac{E_k^2}{k^2} dE(Q,k,a/L).
\end{equation}
A plot of $\Delta E^2$ is shown in
Figure~\ref{figure3}. We wish to emphasize several points: first, at $x=0$
($a=0$), and $x=1$ ($a=L$), the energy fluctuations vanish. This is
as expected for a volume of zero size, and the volume of the whole
box respectively. However, these two points are discontinuous in the
$Q\to\infty$ limit. In the vicinity of these points $dE(x)$ rises
sharply and then remains at an almost constant value. This can be
seen by expanding  in powers of $1/Q$
for $0<x<1$. We find
\[
    dE(Q,k,x)=Q^2+\frac{1}{4}k^2(1-(1-2x)^2)-\frac{k}{\pi^2} + \mathcal{O}(\frac{1}{Q^2}).
\]
If we consider fluctuations for a low energy state (though not the
ground state), we have
\begin{equation}
\label{E:one_d_harmonic_e}
    \Delta E^2=n E_Q^2 + \mathcal{O}(1).
\end{equation}
Where, again, the expansion parameter is $1/Q$. Note that the UV divergence of the
energy fluctuations is linear in the cutoff parameter. Our choice of $\pi^2 Q^2$ as
the cutoff was made to bring eq.(\ref{E:one_d_harmonic_e}) into a convenient form.

In the d-dimensional case, we have:
\[
     N_{mn}=\prod\limits_{i=1}^d \frac{1}{\pi(m_i-n_i)}
        e^{\imath \pi (m_i-n_i) \frac{a}{L}}
        \sin\left(\pi(m_i-n_i)\frac{a}{L}\right),
\]
while
\[
    E_s=\sum_j\frac{\hbar^2 4\pi^2 s_j^2}{2mL^2}.
\]

Now,
\begin{equation}
 \label{ekn}
    \sum_s E_k E_s N_{ks} N_{sk} =
    E_k \frac{\hbar^2 4 \pi^2}{2 m L^2} \frac{1}{\pi^2}
    \sum_{s1} \ldots \sum_{sn} \sum_j \prod\limits_{i=1}^d
    \frac{\sin^2(\pi(k_i-s_i)a/L)s_j^2}{(k_i-s_i)^2}.
\end{equation}
Defining the quantity
$M(s_i)\equiv\frac{\sin^2(\pi(k_i-s_i)a/L)}{(k_i-s_i)^2}$, we can
rearrange the sum in eq.(\ref{ekn}) as follows,
\[
    E_k \frac{2 \hbar^2}{m L^2}
    \sum_{k=1}^d \sum_{s_k} \sum_{j=1}^d \Pi_{i=1}^d M(s_i) s_j^2
   =
    E_k \frac{2 \hbar^2}{m L^2}
    d\left[
        \left(\sum_{s_1} M(s_1) s_1^2 \right)
        \left(\sum_{s_2} M(s_2) \right)^{d-1}
     \right].
\]
Now, according to eq.(\ref{E:Efluc_Harmonic_2}), $\sum_{s_1} M(s_1) s_1^2 = \pi^2
dE(Q,k,x)$ and $\sum_{s_2} M(s_2)$ is the one dimensional $N_{mm}$ operator which
was shown to be equal to $a/L$. Therefore, the energy fluctuations are given by
\begin{equation}
\label{efluctsf}
    \Delta E^2=n d E_k \frac{\hbar^2 4 \pi^2 dE(Q,k,x)}{2 m L^2}
    \left(\frac{a}{L}\right)^{d-1}.
\end{equation}
If $Q$ is large and $a/L$ is small this reduces to
\[
    \Delta E^2=n E_k E_Q d \left(\frac{a}{L}\right)^{d-1} + \mathcal{O}(1),
\]
which has the claimed area scaling behavior \footnote{The term $d(a/L)^{d-1}$ is
suggestive of a sub-leading expression, where the volume term cancels away. This
type of behavior has been discussed in detail in \cite{AreaScaling}.}.

\section{Conclusions and Discussion}

In this paper we calculated fluctuations of the particle number operator and the energy operator restricted to sub-volumes. The whole system
was in a pure state  with fixed particle number and fixed momentum. We discussed systems whose Hamiltonian is time-independent with periodic or Dirichlet boundary conditions. In general the fluctuations differ from their classical counterparts.

Even for a single  non-relativistic particle in a box, the energy fluctuations in a sub-volume with sharp boundaries are divergent and scale with the area of the boundary of the sub-volume. This behaviour depends weakly on the specific state of the system, its boundary conditions, or on the specific Hamiltonian.

In addition to scaling with the area of the boundary the energy fluctuations diverge with the UV cutoff. Technically this cutoff appears when we sum over energy eigenmodes: by introducing a high-momentum cutoff on the Fourier modes in the sum, the energy fluctuations become finite and proportional to the value of the UV cutoff. We have shown that introducing the cutoff is equivalent to smoothing the boundary of the sub-volume such that its width is inversely proportional to the cutoff.

One might have thought that the fact that the divergences are associated with the UV modes suggests that the origin of the divergence requires some additional information on the UV behaviour of the system. We believe that the resolution lies elsewhere: it seems that the divergence of the energy fluctuations is a result of our assumption regarding sharp boundary conditions of the subvolume. Such sharp boundaries would be impossible to construct in practice. This is similar to an attempt to compute momentum fluctuations of a particle which is sharply localized: if the position of the particle is known with infinite precision then the momentum fluctuations formally diverge. Smearing the position of the particle will render the momentum fluctuations finite and inversely proportional to the smearing scale.
It seems as if the divergence in the energy fluctuations is of similar origin, i.e., measuring the energy fluctuations in a volume with an infinitely sharp boundary yields an infinite result. Introducing a high momentum cutoff is equivalent to smearing the boundary of the sub-volume. Put differently, the width of the boundary is inversely dependent on the cutoff scale. A similar effect appears in relativistic field theories \cite{AreaScaling} and it would be interesting to strengthen these suspected ties with the uncertainty principle by performing additional calculations.

For the case of particle number fluctuations we found that the differences between the quantum and classical cases are more subtle. This is related to the fact that number fluctuations do not formally diverge even for an infinitely sharp boundary of the sub-volume. Nevertheless there are differences between the classical fluctuations and the quantum one which depend more strongly on the boundary conditions, the nature of the global quantum state and the Hamiltonian of the system.

\begin{appendix}
\section{Fourier transforms of smoothed step functions}

The basis of eigenfunctions in a one dimensional periodic box of size $L=1$ is
$
\psi_n= e^{2\pi \imath x n}
$
where $n$ is an integer $-\infty <  n < \infty$.
Any periodic function can be expanded in this basis,
\begin{equation}
f(x)=\sum f_n |n\rangle
\end{equation}
and the coefficients are given by
\begin{equation}
f_n=\int_0^1 dx f(x) e^{- 2\pi \imath x n}.
\end{equation}

The (periodic) step function is defined as $H(x-a)=\begin{cases} 1& 0\le x \le a  \\0 &  x > a \end{cases}$ and has the following expansion coefficients
\begin{eqnarray}
H_n &=&  \int_0^1 dx H(x-a) e^{- 2\pi \imath x n}=
\ e^{-\pi \imath a n}\ \frac{\sin[\pi  a n]}{\pi n}.
\end{eqnarray}
The  step function has a useful analytic representation,
\begin{equation}
H(x-a)= a+\frac{1}{2\pi i} \log\left[\frac{1-e^{-2\pi i (a-x)}}{1-e^{+2\pi i (a-x)}}\right]+
\log\left[\frac{1-e^{-2\pi i x}}{1-e^{+2\pi i x}}\right].
\end{equation}
Defining $z=1-e^{- 2\pi i (a-x)}$ and $w=1-e^{- 2\pi i x}$ we observe that logarithms depend only on the ratios $z/\bar{z}$ and $w/\bar{w}$, hence they depend only on $\arg{z}$ and $\arg{w}$.
The steps occur when the branches of the logarithms change abruptly at $x=0$ and $x=a$ when the absolute value of $w$ or $z$ vanish.

The  integral $\int\limits_0^a f(x) dx$ can be treated as a convolution with a step function and hence
\begin{eqnarray}
\label{conv}
\int\limits_0^a f(x) dx &=& \int\limits_0^L dx H(x-a) f(x)= \sum_n H^*_n f_n.
\end{eqnarray}
In eq.(\ref{divdeltae}) appears a sum of the form $\sum\limits_s H_{s-k}^* H_{s-k} E_s E_k$ which results from a product of two step functions and two energy operators.

Introducing a UV cutoff in the sums that appear in eq.(\ref{conv}) or in eq.(\ref{divdeltae}) amounts to suppressing the coefficients $H_n$ beyond a certain large $n_c$, $H_n\to 0$, $n\ge n_c$. An exponential cutoff amounts to replacing $H_n \to H_n e^{-|n|/n_c}$. To see explicitly the effect of an exponential cutoff we may evaluate the position space representation of the cutoff expansion
\begin{equation}
\widetilde{H}(x-a)= \sum H_n e^{-\frac{|n|}{n_c}}=  a+\frac{1}{2\pi i} \log\left[\frac{1-e^{-\frac{1}{n_c}-2\pi i (a-x)}}{1-e^{-\frac{1}{n_c}+2\pi i (a-x)}}\right]+
\log\left[\frac{1-e^{-\frac{1}{n_c}-2\pi i x}}{1-e^{-\frac{1}{n_c}+2\pi i x}}\right].
\end{equation}
The branches of the logarithms now change over a distance $\Delta x\sim 1/\sqrt{n_c}$ near the points $x=0$ and $x=a$. For an interval of size $L$ rather than $1$ the change is over a range $\Delta x\sim L/\sqrt{n_c}$.

We have verified that other cutoff schemes yield qualitatively similar results by performing numerically similar calculations for different cutoff schemes and for different BC.

\end{appendix}

%-----------------------------------------------------------------------------
\begin{acknowledgements}
A.Y. acknowledges the support of the Kreitman foundation fellowship and the
hospitality of the KITP where part of this research has been carried out. We thank
E. Cornell, N. Davidson, M. Einhorn,  D. Oaknin and G. Shlyapnikov for useful discussions and suggestions and Y. Castin, D. Cohen and A. Vardi for comments on the manuscript.
\end{acknowledgements}

%----------------------------------------------------------------------------
%\bibliography{biblist}

\end{document}